\documentclass[pdflatex,sn-mathphys]{sn-jnl}% Math and Physical Sciences Reference Style
\jyear{2022}%
\usepackage{enumitem}
\setlist{nosep} %remove space before itemize environment
\usepackage{caption}
\usepackage{subcaption}
\usepackage{stackengine}

\theoremstyle{thmstyleone}%
\theoremstyle{thmstyletwo}%

\theoremstyle{thmstylethree}%

\begin{document}
\title[CNNCat]{CNNCat: Categorizing high-energy photons in a Compton/Pair Telescope with Convolutional Neural Networks}
%\author*[1,2]{Jan Peter Lommler}
\author*{Jan Peter Lommler$^*$}
\email{lommler@uni-mainz.de}
\author{Uwe Oberlack}
%\equalcont{These authors contributed equally to this work.}
\affil{Institute of Physics \& PRISMA\textsuperscript{+} Excellence Cluster \\
Johannes Gutenberg University Mainz, 55099 Mainz, Germany}
%\affil[2]{PRISMA\textsuperscript{+} Excellence Cluster}

%\begin{abstract}
\abstract{
A Compton/Pair telescope, designed to provide spectral resolved images of cosmic photons from sub-MeV to GeV energies, records a wealth of data in a combination of tracking detector and calorimeter. Onboard event classification can be required to decide on which data to down-link with priority, given limited data-transfer bandwidth. 
Event classification is also the first and one of the most crucial steps in reconstructing  data. Its outcome determines the further handling of the event, i.e., the type of reconstruction (Compton, pair) or, possibly, the decision to discard it. Errors at this stage result in misreconstruction and loss of source information. We present a classification algorithm driven by a Convolutional Neural Network. It provides classification of the type of electromagnetic interaction, based solely on low-level detector data. We introduce the task, describe the architecture and the dataset used, and present the performance of this method in the context of the proposed (e-)ASTROGAM and similar telescopes.

}
\keywords{$\gamma$-Astronomy, Machine Learning, Event Classification, Compton telescope, pair telescope}

\maketitle

\section{Introduction}
%need to add refs to claims
Gamma-ray astronomy in the energy range of sub-MeV to several 10s of MeV remains little explored. At these energies, the Compton effect dominates the total cross-section of photons with matter, changing to dominant pair production in the upper part of this range. The intransparency of the atmosphere requires Compton and pair telescopes to be operated from space.  
Currently, INTEGRAL\,\cite{2003A&A...411L...1W} provides varying sensitivity from 15\,keV to 10\,MeV, while Fermi-LAT\,\cite{Atwood_2009} covers the range from 30\,MeV to 300\,GeV. Higher energy photons are observed by ground-based observatories like H.E.S.S.\,\cite{hess}, Veritas\,\cite{veritas} and MAGIC\,\cite{magic} or, in the future, CTA\,\cite{cta}. More than 20-year old all-sky maps in 0.7 -- 30\,MeV are available from the Compton telescope COMPTEL\,\cite{comptel} on the Compton Gamma-Ray Observatory (CGRO), yet at the limited sensitivity of a pioneering instrument. Comparatively poor sensitivity in the energy range from about 0.1\,MeV to 100\,MeV leaves a window open for discovery, often referred to as the \emph{MeV Gap}.

INTEGRAL as well as Fermi-LAT are successful observatories with a long service record, but at some point they will need to be replaced. In fact, INTEGRAL will be placed in standby at the end of 2024. 
The likelihood for a large-scale mission like CGRO with four gamma-ray instruments to happen again soon is low. This motivates the development of detector concepts that cover the MeV gap and beyond with one instrument at the cost scale of a medium-size mission. A possible solution to this challenge is a Compton/Pair telescope. Example design proposals are the ASTROGAM detector family (e.g., e-ASTROGAM \cite{2017ExA....44...25D}, AllSky-ASTROGAM\,\cite{as-astrogam}) and AMEGO\,\cite{amego}.

\begin{figure}[htb!]
    \centering
    \includegraphics[width=\textwidth]{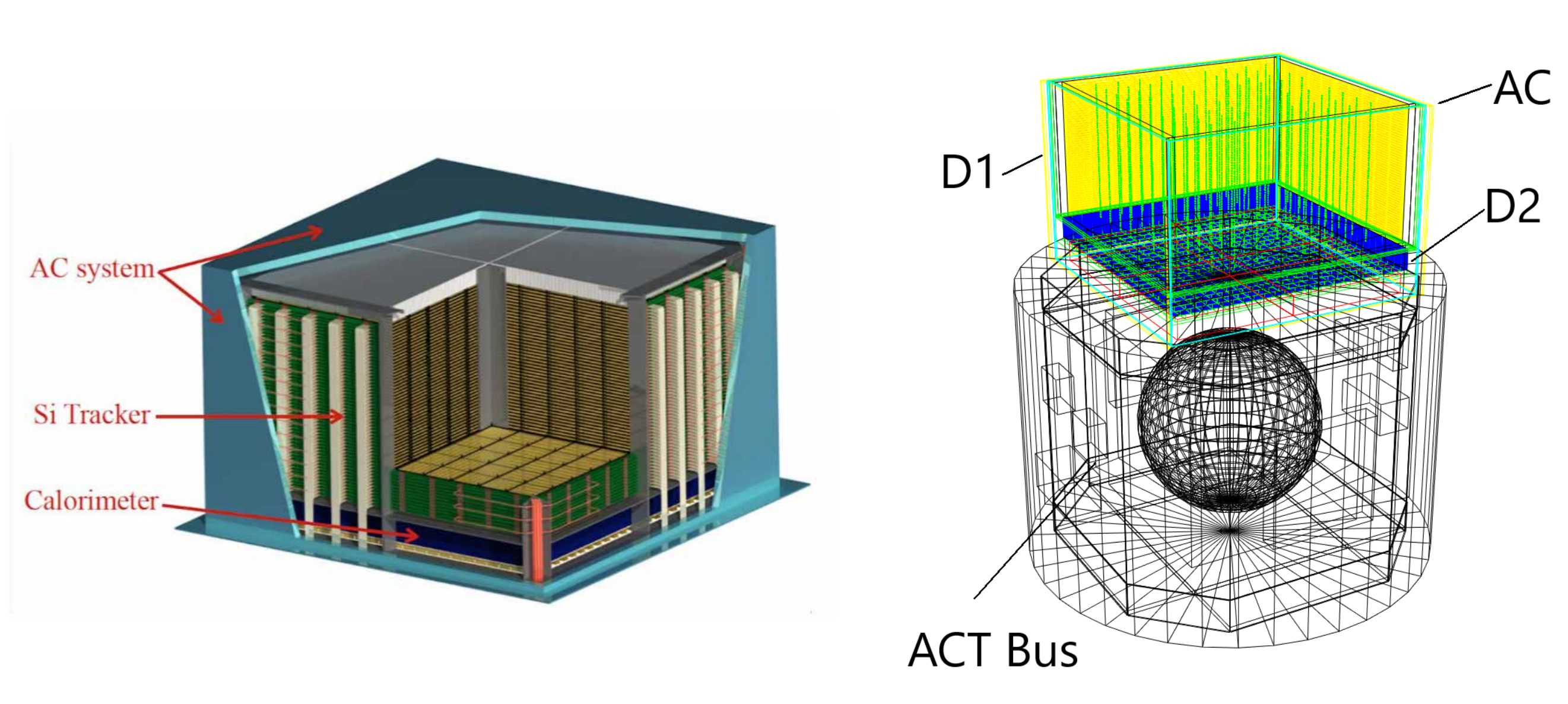}
    \caption{
    Left panel: Schematic view of e-ASTROGAM, consisting of 56 layers of DSSD-tracker (D1) on top of a pixelated CsI(Tl) calorimeter (D2). D1 and D2  are surrounded by an active veto made of plastic scintillator\,\cite{2017ExA....44...25D}. Right panel: Monte Carlo mass model generated with MEGAlib\,\cite{2006NewAR..50..629Z}.}
    \label{fig:CompairSchematics}
    \label{fig:massModel}
\end{figure}

\begin{figure}[htb!]
    \centering
    \includegraphics[width=0.75\textwidth]{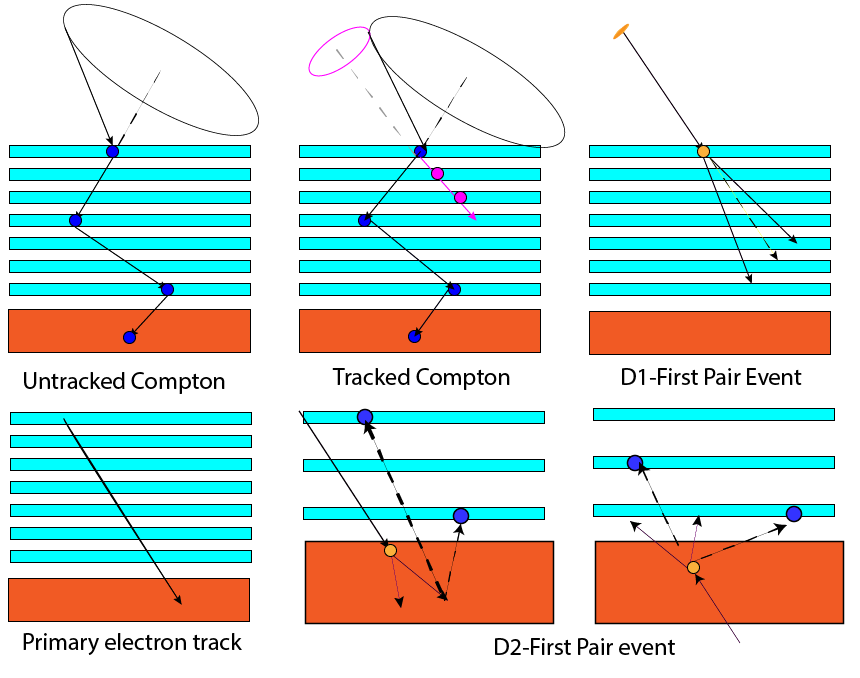}
    \caption{Event topologies encountered in a Compton/pair telescope. Compton scatters can occur without and with a trackable recoil electron (top left and center). To determine the cone of origin of the source photon, the correct sequence of first and second hit has to be identified. For tracked Compton events, the recoil electron track must be isolated from the photon hit pattern and reconstructed separately. This defines a second cone of origin, whose intersection with the photon cone further constrains the origin of the photon. Pair events are reconstructed using the opening angle between electron and positron, providing the photon's origin and energy if the pair-creation vertex occurs in the tracker (top right). Possible background events (bottom row) include charged particle tracks (bottom left) and pair events created in the calorimeter (bottom center and right): Bremsstrahlung, annihilation photons, and electron-positron tracks can mimic legitimate Compton or pair events with no correlation to the original source photon's energy and origin.}
    \label{fig:topologies}
\end{figure}

\subsection{Imaging with a Compton/Pair telescope}
For energies between several hundred keV and several GeV the interaction of $\gamma$-radiation with matter is dominated by Compton scattering and pair production. 
A telescope that is able to image both Compton and pair events, and to determine the energy of the source photon is called a Compton/Pair telescope. It can cover up to five orders of magnitude of the electromagnetic spectrum. 
This study uses the e-ASTROGAM design proposal\,\cite{2017ExA....44...25D} as a reference detector but the concept could be applied to other Compton/Pair telescopes accordingly. Fig.\,\ref{fig:CompairSchematics} shows a schematic of the e-ASTROGAM concept.
The main components are a tracker (D1) consisting of stacked Double-Sided Silicon Strip Detectors (DSSD), a calorimeter unit (D2) made of pixelated Tl-doped cesium iodide (CsI(Tl)), which allows for a depth and x/y-position-sensitive energy measurement, and an active veto system (AC) made of segmented plastic scintillator to prevent triggering on charged primaries. This design principle enables the tracking of photons through their multiple hits in a Compton-scattering sequence, as well as of charged particles as they traverse the detector layers. The ability to accurately measure the energy deposit of charged particles allows the instrument to  distinguish between tracked and untracked Compton events, pair events, and charged-particle tracks originating from, e.g., cosmic rays or albedo secondaries, as depicted in Fig.\,\ref{fig:topologies}.

\subsection{Misclassification and Data Handling in Space}

Bandwidth in Low Earth Orbit is a limited resource. It has to be shared with housekeeping data, satellite telemetry, and secondary instruments on the respective satellite bus. 
One important goal of this study therefore consists in reliably identifying and discarding events that will not be usable in a later stage of event reconstruction. 
Event classification is crucial in reaching this goal. A small and computationally lean model that divides events into potentially reconstructable "good" events and non-reconstructable or background events can help to reduce the amount of data that has to be transmitted. It can also improve event-type-specific preprocessing onboard. 
Application of an incorrect reconstruction algorithm to an otherwise valid event leads to a false reconstruction of source photons into the background. 
The purpose of this study is to investigate whether it is possible to classify events with high efficiency and speed into \emph{Compton events}, reconstructable \emph{Pair events}, and \emph{non-reconstructable events} (3-class classification), while restricting the evaluation of the pursuing improvement to energies up to 20 MeV.

\subsection{Deep Feedforward Convolutional Neural Networks}

A deep feedforward neural net defines a mapping 
\begin{equation}
     \mathbf{y} = f(\mathbf{x}, \mathbf{\Theta})
\end{equation}
where $\mathbf{x}$ is an input vector and $\mathbf{\Theta}$ are the learned model parameters \cite{Goodfellow-et-al-2016}.
In the case of a classification task, $\mathbf{y}$ is identified with a category. Feedforward means that data is processed inside the model only in one direction towards the output nodes. %There are no feedback connections that loop  data back into previous layers of the model.

Convolutional Neural Nets (CNNs)\,\cite{CNNs} emerged as the go-to processing approach when dealing with images and gridded data with local dependencies. The main idea is to restrict the search for relevant information to small regions of the input by applying a moving filter with \textit{learned} kernel values on the pixel values. The number of free (trainable) parameters of a pure CNN is independent of the input size and lower than in any conventional Multi Layered Perceptron (MLP) approach since kernel sizes in general do not exceed $7 \times 7 = 49$ entries. This is much smaller than the actual number of pixels in an image, which would form the input for an MLP. 
The number of calculations necessary in a CNN depends on the size of the input. 
To scale down the processing overhead, a common strategy is  to  sample down the intermediate input by applying either maximum or average pooling. Here a moving window  is applied to the input, propagating either the maximum or the average value of the pixels inside the window. 

CNN-driven classification models are a synthesis of CNN and MLP. The task of the CNN is to extract key features of the image and deliver them to the classifier. The classifier has to comprehend the data presented by the CNN part of the model and decides to which class the sample belongs. In multi-class scenarios, the final output function is a classification vector whose components are generated by the softmax-function:
\begin{equation}
    \text{softmax}(\mathbf{z})_i = \frac{\exp(z_i)}{\sum_j^N \exp(z_j)} = y_{i, \text{pred}}
    \label{eq:softmax}
\end{equation}
Per definition, the sum over all entries equals to 1 and allows a probabilistic interpretation of the model's output.
$y_{i,\text{pred}}$ is therefore the probability ($0 \le y_{i,\text{pred}} \le 1$) of a given sample to belong to the i-th class.
The weight-adaptation is driven by the minimization of the categorical cross-entropy:
\begin{equation}
    CE = - \sum_{i=1}^N y_{i, \text{true}} \log(y_{i,\text{pred}})
    \label{eq:crossEntro}
\end{equation}
where $y_{i, \text{true}}$ is 1 if the sample belongs to class $i$ and 0 else.

\section{Model Architecture and Training}

\subsection{Architecture}
Convolutions can be applied either on 2D images or 3D volumes, which could be stacks of images forming a video or a binned 3D representation of a detector.
A full 3D representation of the tracker module of e-ASTROGAM using only one 8-bit color channel for the deposited energy results in  3D images with a size of around one GB. This would lead to a very large training dataset and intense VRAM usage during training and inference. Therefore a multi-view approach was chosen. For optimal performance, each view is processed in a separate stack of convolutional modules. The result of each stack is then merged and further processed. 
The design of the tracker makes the use of X-Z,  Y-Z and X-Y-projections of an event a natural choice. In addition to the tracker information (one row of pixels per layer), discretized calorimeter information (one row of pixels per centimeter of crystal height) is included at the bottom of the tracker projection. In order to avoid large memory and processing overhead, the X-Y projection was down-sampled by a factor 10.

This reduces the memory footprint of the input data from around 1 GB down to 0.71 MB per event using the same precision for the deposited energy.

\begin{figure}[hbt!]
    \centering
    \includegraphics[width=\textwidth]{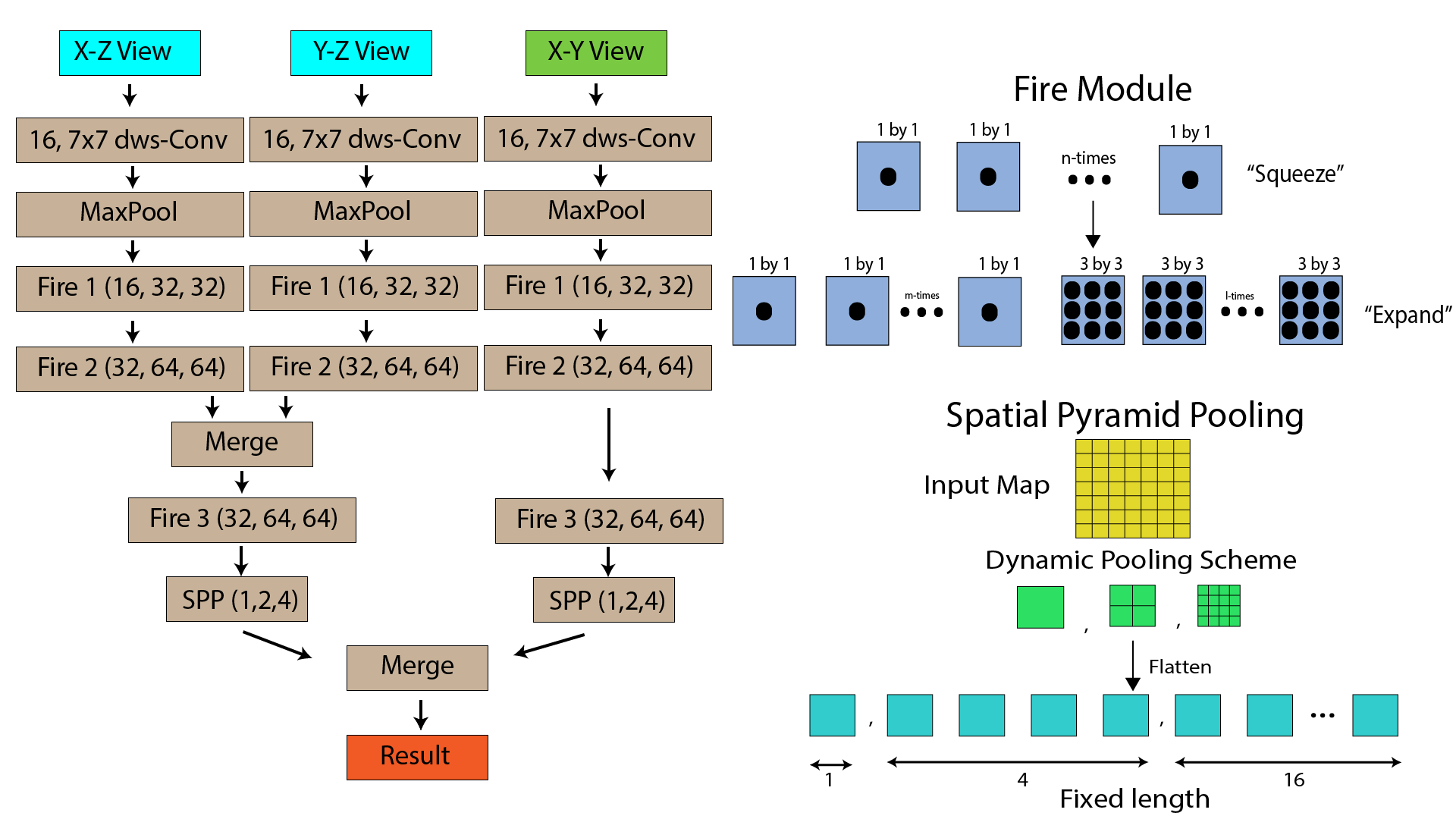}
    \caption{Left: Sketch of the CNNCat model architecture. Each view is processed in a separate stack of convolutional (dws-Conv: depth-wise separable convolution) and pooling layers with identical layout. The results are merged and processed again. The final output is a classification vector whose components give the probability for the shown sample to be either a Compton, Pair or non-reconstructable event. Top Right: Sketch of the micro architecture of a SqueezeNet Fire Module. The number of input maps is reduced to $n$ by the $1 \times 1$ "squeeze" filters that act as information bottle necks. The resulting data is processed by $m$ $1 \times 1$ and $3 \times 3$ "expand" filters. The stacked output is handed to the next layer. Adapted from:\,\cite{DBLP:journals/corr/IandolaMAHDK16}.
    Bottom right: Sketch of spatial pyramid pooling. The incoming feature maps are binned via pooling into $1 \times 1$, $2 \times 2$, $4 \times 4$, ... representations, which are then flattened into the input of the dense layer. Here the length of the outgoing vector is fixed to $n$ $\cdot (1+4+16)$, where $n$ is the number of feature maps to process.  Adapted from:\,\cite{DBLP:journals/corr/HeZR014}} 
    \label{fig:Modellayout}
\end{figure}

%A preliminary investigation using the approach presented by \cite{Aurisano_2016} led to the conclusion that the use of their Inception modules would lead to a rather slow and computationally expensive model.
The SqueezeNet architecture \cite{DBLP:journals/corr/IandolaMAHDK16}, enables the design of small, yet powerful models, which have been demonstrated to run on small-scale hardware like FPGA-accelerated System-On-Chips (SoCs)\,\cite{zynqNet} and was chosen as design pattern of the convolutional modules. The presented model makes heavy use of SqueezeNet's Fire modules shown in the top right of Fig.\,\ref{fig:Modellayout} and is sketched in Fig.\,\ref{fig:Modellayout} left.

A common restriction in Feedforward Neural Nets in general lies in the fixed input size. In the case of CNNs, this restriction is self-imposed by using a dense layer to perform the classification. If used conventionally this layer takes the stack of final feature maps, flattens them, and uses the resulting 1D tensor as input. The number of free parameters introduced by the dense layer for processing is
\begin{equation*}
    %parameters = \textit{filter length} \cdot \textit{filter width} \cdot \textit{filters} 
    \text{parameters} = \text{input length} \times \text{output nodes} \times \text{maps}
\end{equation*}
%example: feature map size 100 by 100 pixels, 7 classes, 32 maps -> 2.2 million parameters
%
and can quickly reach millions because of the size and number of the feature maps to flatten. Due to their nature of connecting every input node with every output node, dense networks correlate each pixel in a feature map with every other pixel, which is also a source of faulty correlations either by pure chance or simulation biases. A more robust approach called Spatial Pyramid Pooling was introduced by\,\cite{DBLP:journals/corr/HeZR014}. 
The final feature maps are binned by applying Pooling with a set of dynamically sized windows. When flattened, this results in a fixed input size for the dense layer, regardless of the original image size. Each value now corresponds to regions of the processed feature map, effectively binning the image into $n \times n$ equally sized rectangles. 
This facilitates decoupling of the input size of the classifier (the dense layer) from the output of the feature generator (convolutional neural net). In addition, the size of the input vector of the dense layer is generally smaller than with conventional flattening. This reduces the number of free parameters. He et al.\,\cite{DBLP:journals/corr/HeZR014} also demonstrated that contemporary image classification networks using this approach outperform the designs with conventional flattening. In this study, max-pooling was used to generate the binned representation. Another application of this pooling scheme is to resize intermediate input with respect to each other. The feature maps of the X-Z filter tree have a different size with respect to the X-Z and Y-Z maps. This can be ramified by applying the dynamical pooling scheme in order to generate compatible input for the dense layer.

One go-to method for regularizing neural networks (i.e., to prevent overfitting) is Dropout\,\cite{Dropout}. During training in layers with Dropout applied (usually the dense layer or the first few convolutional layers), randomly selected neurons are switched off for each training step. To compensate, the model has to learn to propagate vital information for the classification over various paths in the graph. During inference, no Dropout is applied, i.e., the full model is used. Using weight adaptation only during training results in a more robust model.
 
The CNNCat model was implemented using TensorFlow\,\cite{tensorflow2015-whitepaper} Version 2 with TensorFlow's internal keras implementation. Spatial Pyramid Pooling was implemented with a modified version of\,\cite{kerasSPP}.

\subsection{Dataset Generation and Selection Criteria }
The training and test data generation is based on MEGAlib\,\cite{2006NewAR..50..629Z} simulations with the mass model shown on the right in Fig.\,\ref{fig:massModel}. The simulation output was then processed into X-Y, X-Z and Y-Z images using custom-written python scripts. Each Z-projection has a size of 66 rows by 4267 columns with 3 color channels. The X-Y projection has a size of 427 by 427 and also 3 channels. Channel 1 contains the binned deposited energy in the pixel saturating at 2048 keV, channel 2 contains the deposited energy relative to the total energy deposited in the tracker, while channel 3 contains the deposited energy relative to the total energy measured in the calorimeter. If the calorimeter has no deposited energy, the values of channel 3 is set to 0.   Since the event topology changes with incident angle and energy, the simulations were split as shown in Table\,\ref{tab:trainSetComposition}. This allowed for a controlled dataset composition decoupled as much as possible from detector efficiencies.
\begin{table}[hbt!]
    \centering
    \caption{Data binning used for dataset balancing and simulation management.  }
    \begin{tabular}{c|c|c|c|c}
        \hline
        Type &  Low energy & Medium energy & High energy & $\Theta$\\
        \hline
        Compton & 0.1 - 1 MeV & 1 - 5 MeV &  5-10 , 10-15, $>$15 & 0-180   \\
        \hline
        Pair &10-200 MeV & 0.2 - 1 GeV & 1 - 3.5 GeV & 0-180  \\
        \hline
        Electron & 10 - 100 MeV & 0.1 - 1 GeV & 1 - 10 GeV &90-180\\
        \hline
    \end{tabular}
    
    \label{tab:trainSetComposition}
\end{table}

\begin{figure}[hbt!]
    \centering
    \includegraphics[width=\textwidth]{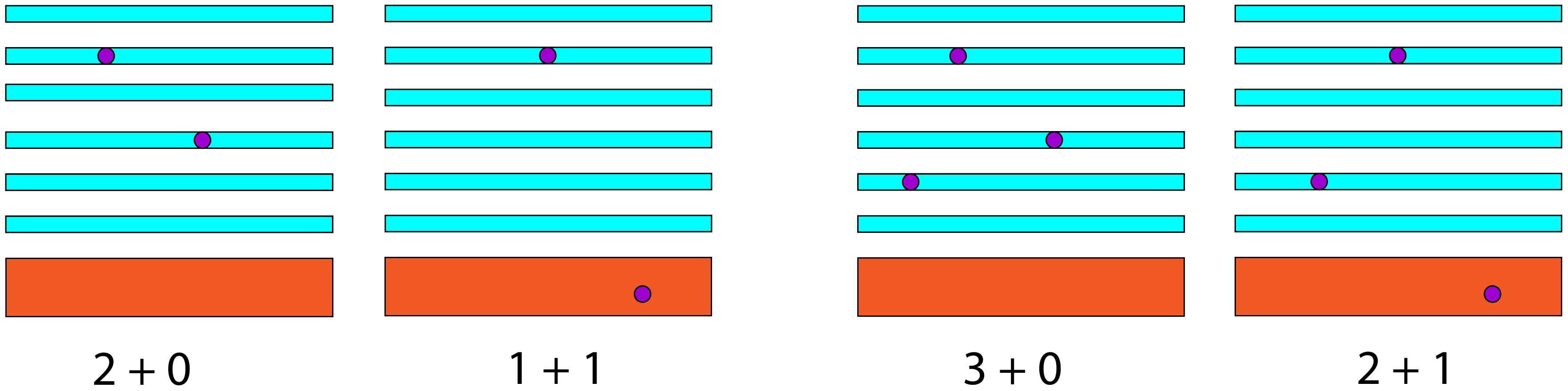}
    \caption{ Enforced trigger pattern to ensure a minimum amount of reconstructable information in each event. Compton events have to fulfill at least 1+1 or 2+0, pair events have to fulfill at least 3+0 or 2+1.}
    \label{fig:trigger:pattern}
\end{figure}

In addition to enforcing the absence of a veto trigger, we restrict the training and test data to events that can be reconstructed in principle, as shown in Fig.\,\ref{fig:trigger:pattern}. In order to reconstruct a Compton sequence one needs at least two hits. Since the position and energy resolution in the calorimeter is lower in general, a coincidence of at least two individual tracker layers or one tracker layer and the calorimeter is required for Compton events included in the training set. Pair events require at least three resolvable hits.

In this study, we use the following definitions of event types:
\begin{itemize}
    \item \emph{Compton event:} The first interaction in the sensitive part of the detector is a Compton scatter and the veto was not triggered (Fig. \ref{fig:topologies}, top left and center).
    \item \emph{Pair event:} Electron and positron interacted inside the sensitive part of D1 first and the veto was not triggered (Fig. \ref{fig:topologies}, top right)
    \item \emph{Non-reconstructable:} Events that cannot be reconstructed meaningfully. Either the event is resulting from an electron primary traversing the geometry without triggering the veto, or a pair was created inside D2 and did not trigger the veto. D2-first pairs are a case of lost source photons, events triggered by electron primaries are a genuine background component (Fig.\,\ref{fig:topologies}, bottom row).
\end{itemize}

It was also deemed important to enforce clean samples from which the model is tasked to learn the characteristics of good Compton and Pair events. The very first interaction of a photon in  each training set sample, independent of the type of event, has to occur in sensitive detector material. Pair events have to be created inside the tracker.
Since vetoed events in space-borne detectors are usually discarded immediately,  they are  not considered in the datasets of this study.
For image augmentation, we applied Gaussian noise with the estimated energy resolution of the proposed detector to the input images to further solidify the model against overfitting.\footnote{Since the D1 detector is highly inhomogeneous due to its layered design, the common technique of rotating the image cannot be applied.}

\section{Results}
We start the presentation of the results with an overview of the computational demands, introduce performance measures used throughout the evaluation, followed by a performance analysis on two test sets. Test set 1 comprises the full observational range from 100\,keV to 3.5\,GeV from all incidence angles. Test set 2 considers a simulated  Crab observation.

\subsection{Computational Demands and Architectural Optimizations}
Since one of the possible applications should be an online data classifier used on the satellite itself, all the design choices were driven toward optimizing the number of computations per event.
The choice of Fire-Modules as micro-architecture for the convolutional modules allows us to control the size and hence the computational demands of the model, due to their built-in bottlenecks.
Dropout as a regularization method introduces no computational overhead at run-time while effectively constraining the model and preventing overfitting of the training data. 

To further optimize the number of computations per convolutional layer, depth-wise separable convolution (DWS-Conv)\,\cite{SifreDepthwiseSep} was applied wherever feasible, i.e., on the $7 \times 7$ and $3 \times 3$ convolutions\footnote{This speed-up may not be realized with non-optimized TensorFlow models since the underlying libraries (e.g. CuDNN) may not be optimized for this convolution type. Running the model through an inference engine like tensorRT offered by NVIDIA will show the expected behaviour.}.

A conventional convolutional layer applies $m$ $n\times n$ convolutions on $l$ incoming channels on an image with $j\times k$ pixels, where $m$ denotes the number of filters to be applied and $n$ is the filter size along one axis in pixels.  Assuming $m = 16$, $n=3$, $l=3$ and $j=k=100$, this setup would result in 
\begin{equation}
    \text{parameters} = n^2 \times l \times m = 432 
\end{equation}
and 
\begin{equation}
    \text{ops} = j \times k \times n^2 \times l \times m = 7 \times 10^6
\end{equation}
operations used in this layer. 

Depth-wise separable convolution, on the other hand, tries to simplify this step without sacrificing too much performance. In this scheme, only one full $n \times n$ convolution is applied to each incoming channel. The intermediate data is put back together and then stretched to the required dimension by applying $m$ $1 \times 1$ convolutions. 
In the setup described above, this results in 
\begin{equation}
    \text{parameters} = n^2 \times l + l \times m = 75
\end{equation}
and 
\begin{equation}
    \text{ops} = n^2 \times l \times j\times k +  l \times j\times k \times l \times m = 1.2\times 10^6.
\end{equation}
Applying this to the precursor model allowed us to reduce the number of trainable parameters to 64k and the number of FLOPS per event to $0.9$\,GFLOPS. ESA identified the necessity to explore systems for onboard processing and funded the GPU4S (GPUs for Space) project that surveyed use-cases and possible embedded systems\,\cite{GPU4S}. Follow-up studies also investigated potential improvements of commercial off-the-shelf embedded systems (e.g.\,\cite{GPUFollowUp}). Ortiz et al. also studied the Jetson AGX system which delivers around 1.3 TFLOPS per Watt at FP16 or 2.6 TOPS at INT8, equivalent to a rate of 1.4\,kHz or 2.9\, kHz  per Watt, respectively. Note that either a smaller detector or a coarser binning of x-y strips in the input data would increase the amount of rate per Watt significantly as the computational complexity for CNNs scales mostly with $j \times k$ of the input data.
A more recent study of Rodriguez-Ferrandez et al in cooperation with ESA was able to show that this kind of devices can sustain the radiation environment in orbit \cite{XavierProton}. They obtained acceptable error rates, making them a candidate system for embedded systems especially designed for space applications. 
%According to NVIDIA's claims, space-graded systems are to be released. Therefore, a model similar to the one presented here is suited to run on soon to be available embedded systems.

\subsection{Performance Measures}

The outcome of each binary event classification falls into one of the following four cases:
\begin{enumerate}[label=\roman*.]
    \item[(tp)]True positive: the sample belongs to the positive class and is classified correctly (positive).
    \item[(tn)] True negative: the sample belongs to the negative class and is classified correctly (negative).
    \item[(fn)] False negative: the sample belongs to the positive class and is classified as negative.
    \item[(fp)] False positive:  the sample belongs to the negative class and is classified as positive.
\end{enumerate}

From this, several metrics can be derived, each bound between 0 and 1, with 1 being a perfect performance under the given metric. The most common one is \emph{accuracy}, defined as 
\begin{equation}
    \text{Accuracy} = \frac{tp + tn}{tp +tn +fn +fp}.
\end{equation}
It gives the fraction of correct predictions relative to all predictions. Accuracy is a workable metric on class-wise balanced datasets but may give deceptive results when the data is skewed towards one class. 
%For example, consider a case where the signal-to-background ratio is expected to be very small. A (useless) 'classifier' that would tag all events as background would reach an accuracy of almost 100\% while failing the given task completely. 
On unbalanced datasets, performance must be measured with more robust metrics. Two of those, from the context of binary classification and information retrieval, are precision and recall.
\emph{Precision} is defined as
\begin{equation}
    \text{Precision} = \frac{tp}{tp + fp}
\end{equation}
 and is a measure of how reliable a classification into a certain class is, i.e., it is a measure of the purity of the events in that class. It can be influenced by the classifier's performance on the samples of a class itself or by misclassification of samples belonging to other classes. 
 
The other metric is the \emph{recall} on a given class. It is defined as:
 \begin{equation}
     \text{Recall} = \frac{tp}{tp + fn}.
 \end{equation}
Recall, also called true positive rate, is a measure of how well a given class is recognized by a classifier, i.e., the efficiency with which these events are properly classified. 
In general, it is impossible to optimize both precision and recall to arbitrarily high values unless the samples in the dataset are perfectly separable.
%Usually, precision and recall reach an equilibrium state where improvements in one metric imply performance loss in the other.
Both metrics can be extended from binary to multi-class scenarios by applying a one-against-all scheme. The class to be evaluated is considered the positive class, while the other classes are combined into the negative class. False positives then have to be defined as the combined misclassifications into the investigated class from all other classes. 

%\begin{table}[hbt!]
%    \centering
%    \caption{Layout of the confusion matrix for the 3-class task in this study. True positives align on the main diagonal. Each column corresponds to all samples predicted into a given class, each row provides the classification result for a given (true) class. C is for Compton event, P is for pair event, NR is for non-reconstructable event. Every off-diagonal matrix element can be interpreted as either a false negative or a false positive of a given class.  }
%     \label{tab:convMatrix}
    
 %       \begin{tabular}{c|c|c|c}
        
  %         Truth$\backslash$Classification  &  Compton & Pair & non-reconstructable \\
  %         \hline
  %          Compton & tp C & fn C, fp P & fn C, fp NR  \\
  %          Pair & fn P, fp C & tp P & fn P, fp NR\\
  %          non-reconstructable & fn NR, fp C & fn NR., fp P & tp NR\\
  %      \end{tabular}
   
%\end{table}

\subsection{Training and Validation}

\begin{figure}[hbt!]
    \centering
    \includegraphics[width=0.95\textwidth,clip]{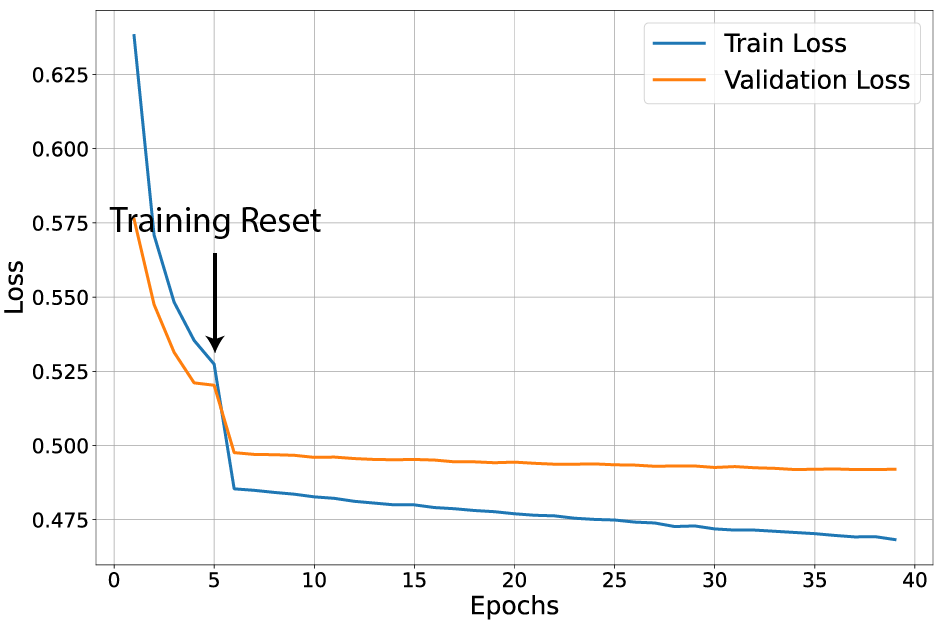}
    \caption{Training Loss as function of epochs. Training was started with Learn Rate I. After epoch 7 early stopping terminated training run 1. The model was reset to the previous best weights. The training was then restarted with Learn Rate II for the remaining 34 epochs.  }
    \label{fig:trainLoss}
\end{figure}

The training set consisted of 400,000 events evenly divided over classes, incident energy and incident angle binning. 300k stratified events were used for training, 100k similar selected events were held back for validation for the purpose of monitoring the training progress. We used the Adam optimizer \cite{kingma2017adam} for parameter adaptation. Early stopping was used to prevent overfitting due to excessive training. During trials we observed that the improvement stalled after only a couple of epochs. To mitigate this, early stopping was used to trigger a restart of the optimizer with a lower learn-rate on the best-performing model (denoted as Training Reset) in order to force the optimizer to a lower minimum. The progress during training is shown in Figure\,\ref{fig:trainLoss}.  The relevant training and hyper parameters are summarized in Table\,\ref{tab:trainSetup}

\begin{table}[hbt!]
    \centering
    \caption{Best training parameters found after optimization. Delta gives the minimum improvement from the previous best loss-value, patience is the number of epochs in which the improvement has to be achieved.  
     }
    \begin{tabular}{c|c|c|c|c|c|c}
     Learn Rate I & Learn Rate II &  Dropout & Batch Size & SPP & Delta & Patience \\
                \hline
         10$^{-4}$ & 10$^{-6}$ & 0.3& 1 &[1,2,4] & 0.001 & 5 \\
    \end{tabular}
    
    \label{tab:trainSetup}
\end{table}

Training for the best-performing model took 39 epochs with a total duration of around 9 days on an NVIDIA GeForce RTX 2080\,Ti graphics card.  The model was reset after early stopping to the best-performing weights after epoch 34, which were then used for evaluation against the test sets.

\subsection{Classification Performance}

The test set contains 2,562,406 events containing photons and electrons ranging from 100\,keV to 3.5\,GeV simulated from a flat spectrum uniformly from all incident angles. The energy range was divided analogously to the training-set generation. 
This dataset is reserved to analyze the energy dependency of the classification result averaged over all incident angle bins and was not used during training or validation.

\begin{figure}[p]
    \centering
    \begin{subcaptiongroup}
    \phantomcaption\label{fig:comptonConf}
    \stackinset{l}{-8pt}{b}{5pt}{\captiontext*{}}{
    \includegraphics[width=0.66\textwidth,trim=75 55 100 80, clip]{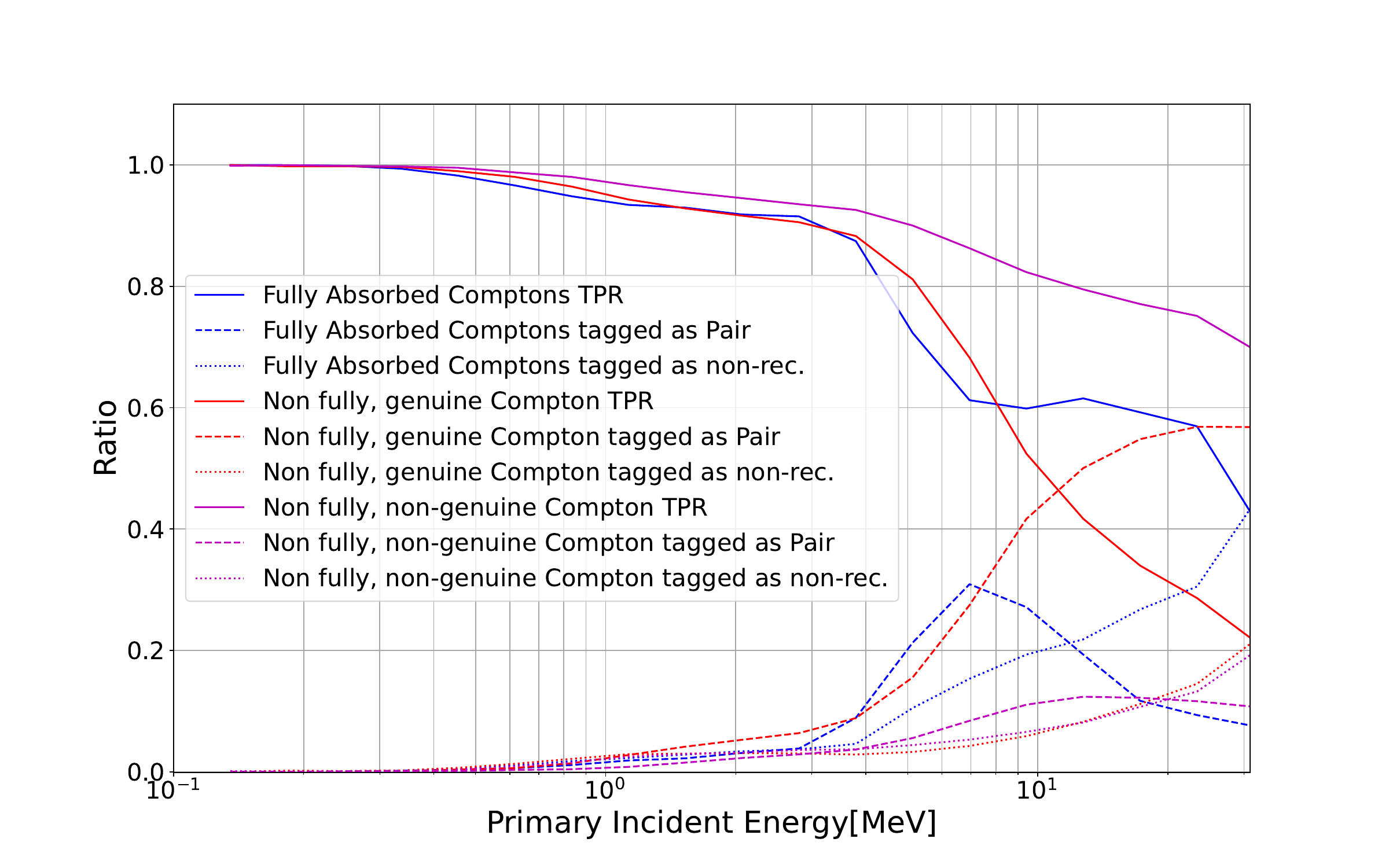}}\\
    \phantomcaption\label{fig:pairConf}
    \stackinset{l}{-8pt}{b}{5pt}{\captiontext*{}}{
    \includegraphics[width=0.66\textwidth,trim=75 55 100 80, clip]{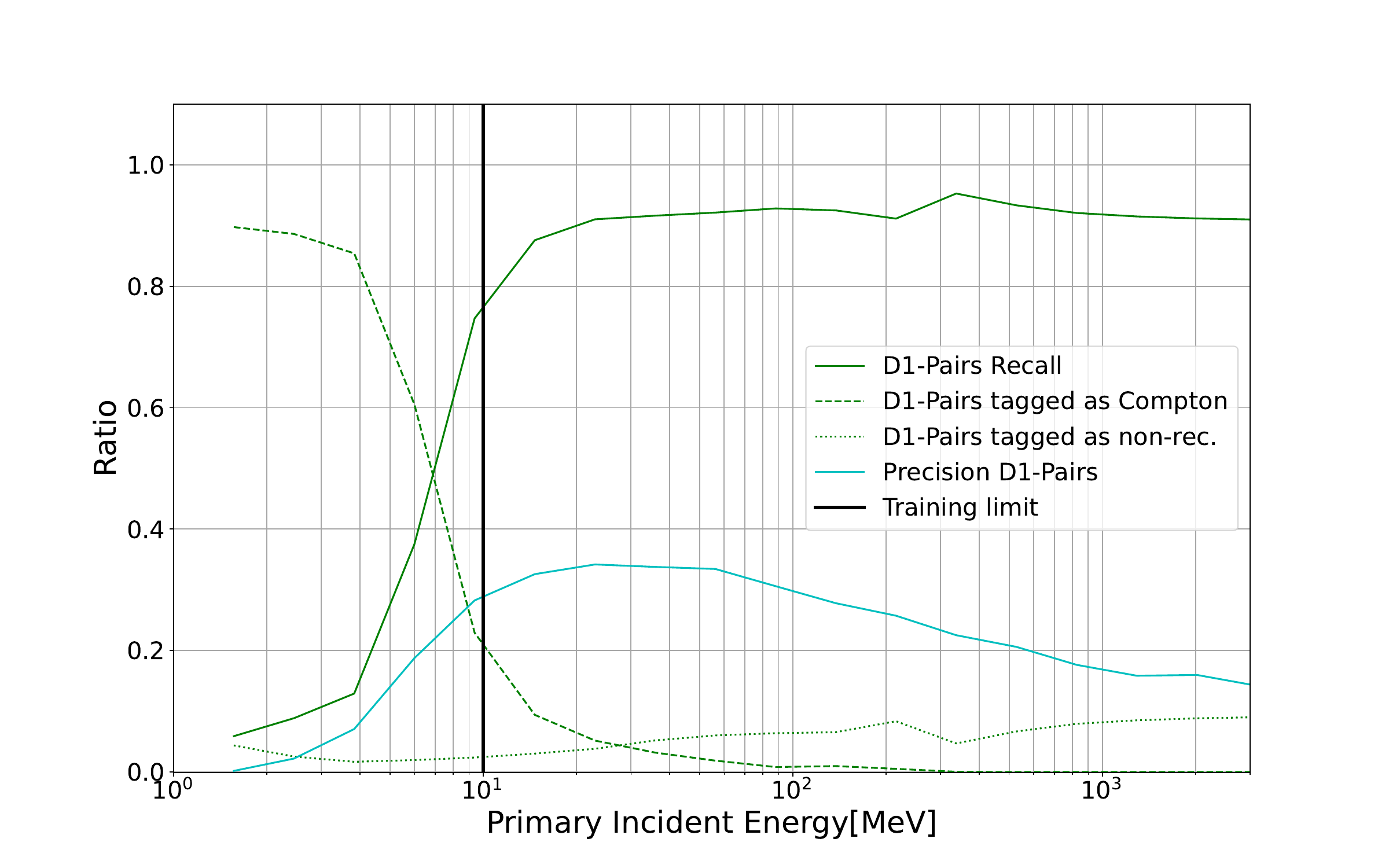}}\\
    \phantomcaption\label{fig:trashConf}
    \stackinset{l}{-8pt}{b}{14pt}{\captiontext*{}}{
    \includegraphics[width=0.66\textwidth,trim=75 25 100 80, clip]{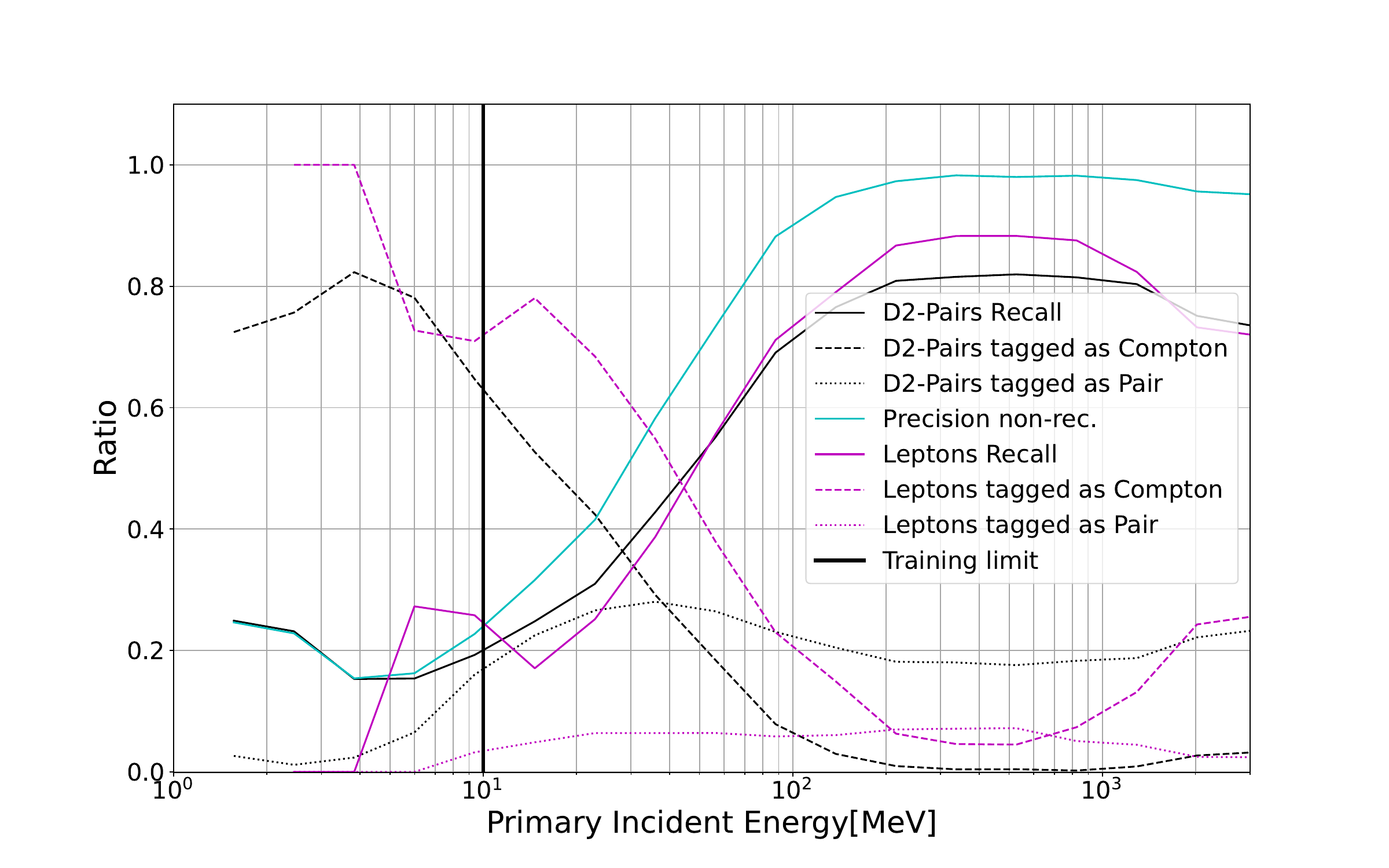}}\\
    \end{subcaptiongroup}
    \caption{
    Precision, recall and class-wise false negative rates vs.\ incident energy.
    \\
    \subref{fig:comptonConf} Compton events. 
    Recall for the three sub-categories of Compton events and corresponding false negative tags vs.\ incident energy up to 30\,MeV. 
    \\
    \subref{fig:pairConf} D1-first pair events. 
    The model was trained on events with $E_{inc} >$10\,MeV, denoted by the bar.
    \\
    \subref{fig:trashConf} Non-reconstructable events.  
    Misclassified events are mostly tagged as Compton for $E<40$\,MeV and as Pair above. 
    The model was trained on events with $E_{inc}>$10\,MeV, denoted by the bar.
    }

\end{figure}

%%%%%%
We evaluate the performance of CNNCat over the full observation range of the instrument using the average over all directions as a function of incident energy of recall, precision, and class-wise false negative rates. The resulting plots are energy-dependent representations of rows of the confusion matrix, shown in Figures \ref{fig:comptonConf} -- \ref{fig:trashConf}. On the input side, we further distinguish Compton events of three types: fully absorbed (golden) events, non-fully absorbed events starting in sensitive material (``genuine'' Compton events - potentially reconstructable), and non-fully absorbed events starting in passive material ("non-genuine" Compton events), which, effectively, are source photons turned into background, as the initial scatter angle and energy loss is unknown.

Fig.\,\ref{fig:comptonConf} shows the model performance on Compton events. The recall of all Compton subsets is very high (above 90\,\%) for incident energies up to 4 MeV. The recall of non-fully absorbed genuine Compton events steadily declines over the remaining energy range, while it stabilizes at 60\,\% at incident energies between 7 to 20 MeV for fully absorbed Compton events . Non-fully non genuine event are less accurately tagged with increasing incident energy but stay above 70\,\% over the shown energy range.    Mistagged fully absorbed Compton events are mostly classified as Pair event at energies around 3 to 12\,MeV. Beyond that, the majority of false negatives is tagged as non-reconstructable. The vast majority of false negative non-fully absorbed genuine Compton events are  wrongly tagged as Pair. This can be explained by the very narrow scatter angle distribution at higher energies, which will give photon hits a topology similar to  tracks which together with a recoil electron track can confuse the model. Another possibility is the generation of multiple recoil electron tracks before  the photon leaving the detector. Non-genuine non-fully absorbed Compton events show a behaviour similar to fully absorbed Compton events, but shifted to higher energies. Here the turnover from Pair to non-rec. tagging occurs around 18\,MeV. 

Fig.\,\ref{fig:pairConf} shows the model performance on pair events with their first interaction measured in D1.  The recall on pair events at incident energies below 7\,MeV is very low. This is not surprising since the model only had pair events with incident energies larger than 10\,MeV in the training data. Electrons and positrons in this energy range are very susceptible to multiple scattering in solid material and show a topology similar to recoil electrons in Compton events. Also, positrons are generating shorter tracks, leading to even greater similarities between Compton and pair events. At 10\,MeV, however, the recall is already passing 78\%. For energies higher than 100\,MeV the recall saturates at 92\%. Precision is in general low due to the large amount of false positives from Compton events at low energies and D2-first pair events over the whole energy range. 

Fig.\,\ref{fig:trashConf} shows the model performance on non-reconstructable events. The recall on D2-first pair events is rising to around 80\% at around 100\,MeV leptonic events are recognized reliably from around 70\,MeV and higher with a peak recall of 92\% around 200\,MeV. In the energy range up to 50\,MeV, the majority of misclassified events is tagged as Compton. Over the remaining range, most of the misclassified events are tagged as pair. Overall precision for energies higher than 100\,MeV peaks at 92\%, leaving  a relatively small number of potentially reconstructable events that were falsely rejected.

\begin{figure}[hbt!]
    \centering
    \includegraphics[width=0.95\textwidth,trim={40pt 10pt 60pt 50pt},clip]{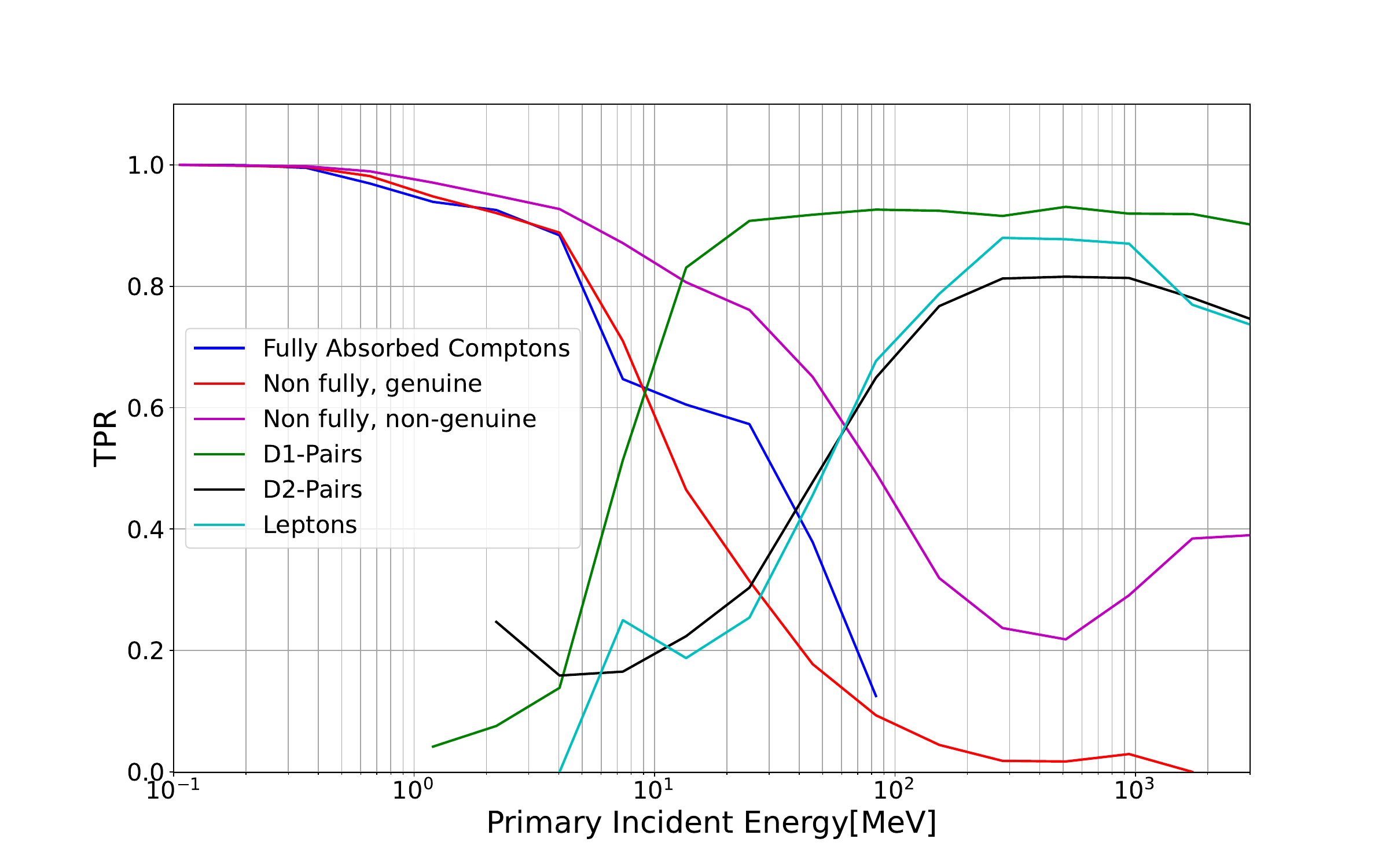}
    \caption{Recall vs. incident energy for the three Compton event populations, D1- and D2-first pairs, and lepton events.}
    \label{fig:totalRecall}
\end{figure}

Fig. \ref{fig:totalRecall} summarizes the recall vs.\ incident energy for all classes, distinguishing the three Compton event populations, D1- and D2-first pairs, and lepton events.

\begin{figure}[hbt!]
    \centering
    \includegraphics[width=0.95\textwidth,trim={40pt 10pt 60pt 50pt},clip]{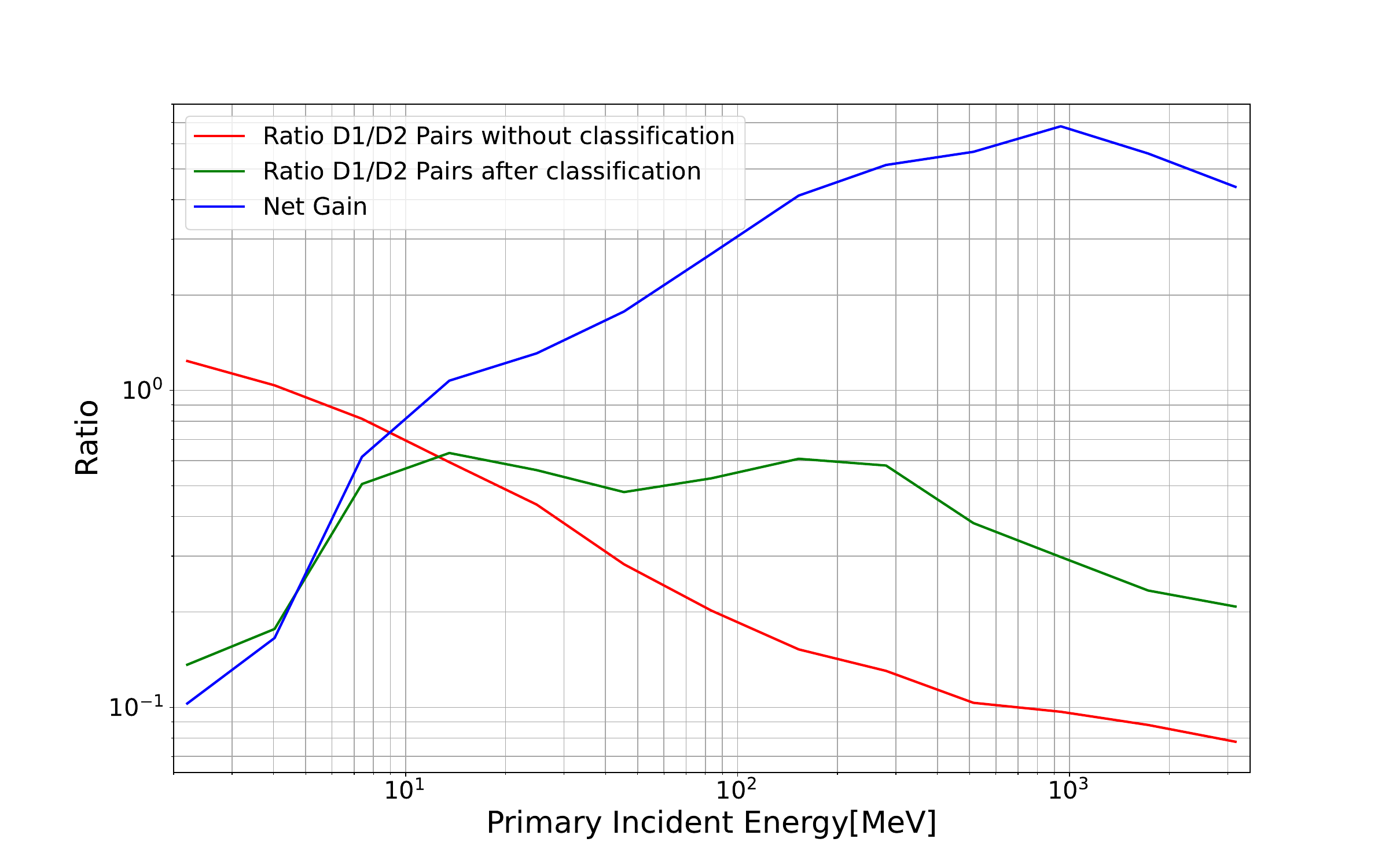}
    \caption{Ratio of D1-first to D2-first pair events vs.\ incident photon energy. At lower energies (below 15\,MeV) the effect of the classification on the ratio is negative. This is because of the lower recall of the model on pair events in this energy region. For higher energies, the model is able to improve the ratio of D1-first vs.\ D2-first pairs by a factor ranging up to 8 at 1000\,MeV.}
    \label{fig:3ClassTrashRatio}
\end{figure}

 Fig. \ref{fig:3ClassTrashRatio} shows the ratio of D1-first ('good) to D2-first ('bad') pairs over the observed energy range. Since e-ASTROGAM was designed to measure both Compton and pair events, the detector design had to compromise between the demands of both detection channels. For instance, the tracker layers lack tungsten converter foil, which enhances the creation of pairs in Fermi-LAT, since this would severely hamper the measurement of (low-energy) Compton events. As a result, many high-energy source photons traverse the tracker without interaction and create pairs only in the calorimeter as indicated by the ratio before classification. This skews the ratio of D1-first to D2-pairs heavily towards non-reconstructable events that are potentially dangerous.  While initially above 1, the ratio soon drops to values below 0.1, manifesting the large surplus of D2-first pairs. This ratio greatly improves after classification for incident energies greater than 15\,MeV. Above 20\,MeV, i.e., in the intended observation range in pair-imaging mode, the model greatly improves the 'good'/'bad' pair ratio by up to a factor 8, shown by the blue line. The weak performance at energies below 10\,MeV is tolerable, since, at these low energies, pair events are difficult to reconstruct and are dominated by Compton events.

Overall, one can conclude from this test set that a small-scale model is capable of distinguishing the presented events by type. As expected, the decision to classify an event as either Compton or pair is biased to some degree by the energy the model can see in an event. This can be seen from the fact that almost no events with an incident energy below 1\,MeV are classified as pair event.  The classification has a beneficial effect on the ratio of D1-first pairs to D2-first pairs for all but the lowest energies.

\begin{figure}[hbt!]
    \centering
    \includegraphics[width=0.95\textwidth,trim={40pt 10pt 60pt 50pt},clip]{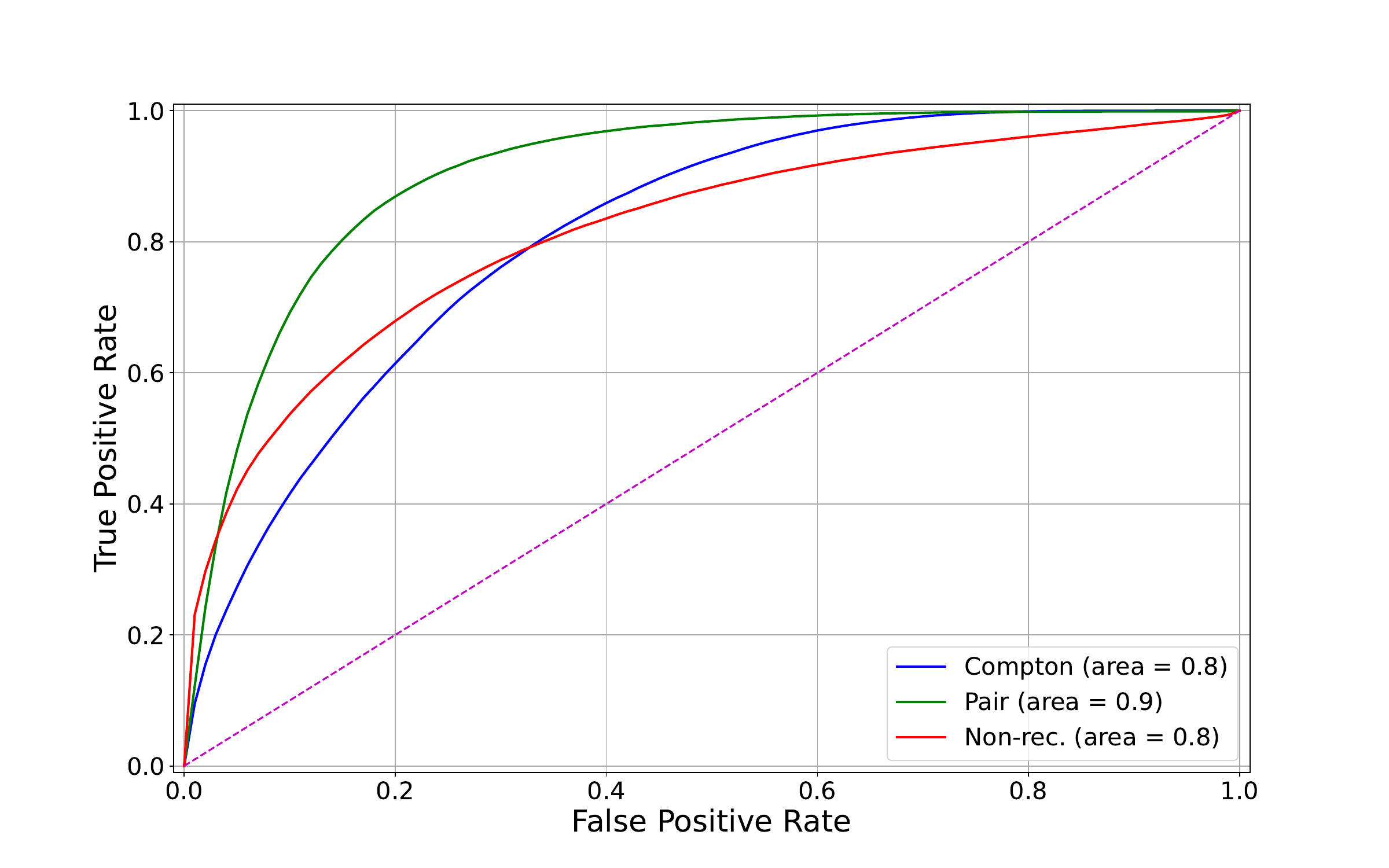}
    \caption{Directionally averaged ROC curves for each class for the test set. The performance of a random classifier corresponds to the dashed line. The Area Under the Curves are given in the legend. Classification of pair events performs best, with an AUC-score of 0.9. The other classes are still performing very well as indicated by scores larger than 0.8.}
    \label{fig:roc_auc}
\end{figure}

To assess the total performance of our classifier, we calculated the Receiver-Operator-Characteristic Area-under-the Curve (ROC-AUC) 
by determining the True Positive Rate and False Positive Rate for decreasing decision thresholds for the positive class. 
In multi-class scenarios this is done in a so called `one vs. all' approach. 
%There is only one positive class and the other classes are merged into a single negative class.  
%This score can be used to assess the performance of a classifier. 
The resulting curves for the test dataset are shown in Fig.\,\ref{fig:roc_auc}. All three classes are well separated from random guessing (indicated by the dashed line). 
There is no task-independent interpretation of the AUC-score, but one can employ a rule of thumb found in \cite{rocAuc}. 
%Performance could be considered `poor' for a score below 0.7, `acceptable' for 0.7 to 0.8, `excellent' for 0.8 to 0.9, and `outstanding' for scores larger than 0.9. 
The scores for each class are around a value of 0.8, which indicates a good performance on the presented dataset. The best performance is achieved by the pair classification, which achieves an area of 0.9. 
%According to the cited performance standards, our model reaches the second highest performance category.
Following the scheme of \cite{rocAuc}, our model reaches the second highest performance category.

\subsection{Crab Nebula Simulation}

\begin{figure}[hbt!]
    \centering
    \includegraphics[width=0.95\textwidth,trim={40pt 10pt 60pt 50pt},clip]
    {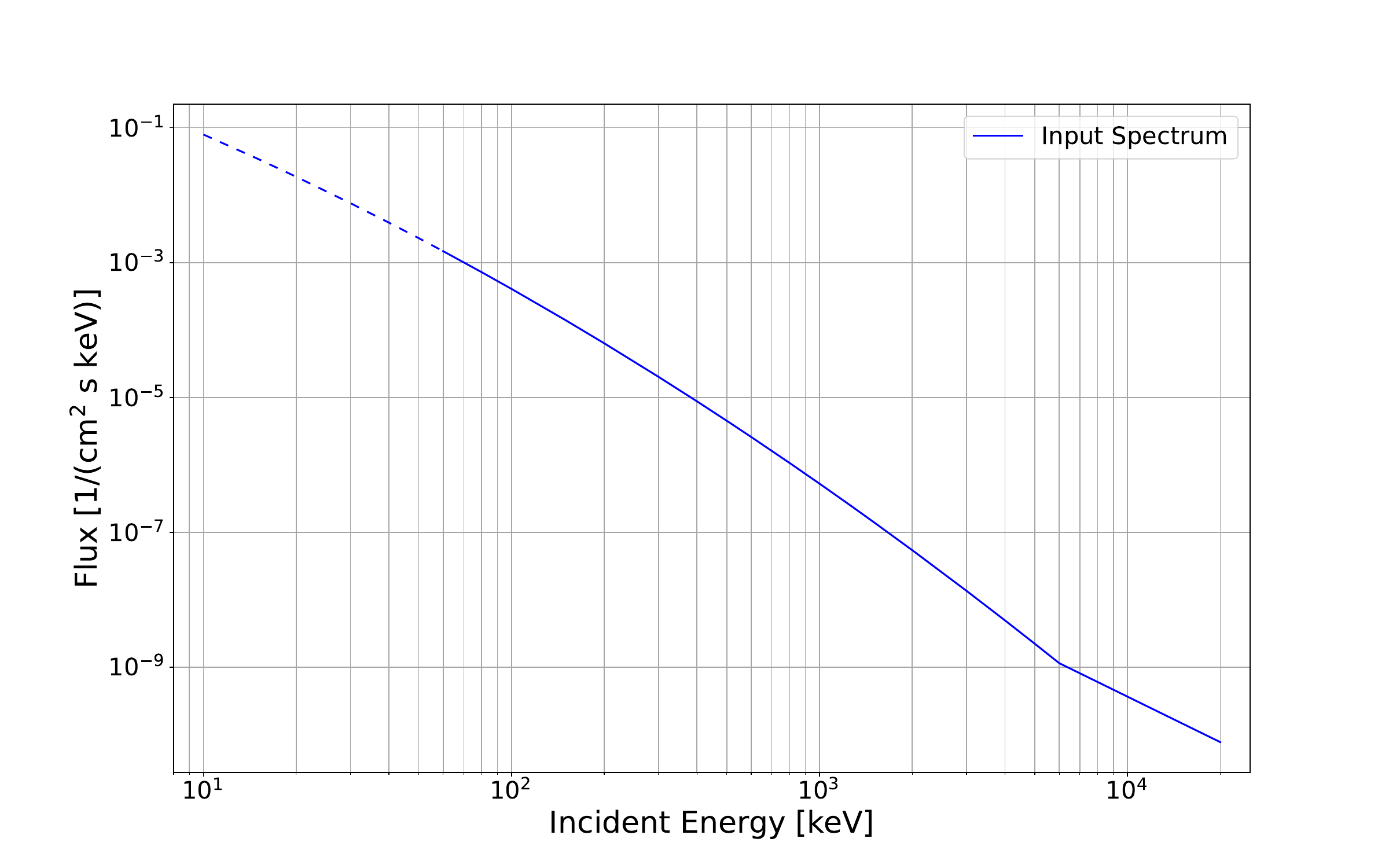}
    \caption{Input spectra for the Crab-Nebula dataset. The Crab spectrum uses ref.\,\cite{crab_spectrum} at $E<6$\,MeV and a power-law with index 2.24 above. The simulation range is indicated by the solid line.} 
    \label{fig:InputSpectra}
\end{figure}

%In this part, we investigate the impact of prior classification on the reconstruction of Compton event data. 
In this part, we investigate the impact of prior classification on the reconstruction of event data up to 20 MeV, applying the standard MEGAlib reconstruction chain on a simulated Crab Nebula measurement.
The Crab spectrum was modelled according to ref.\,\cite{crab_spectrum} for energies up to 6\,MeV, and extended to 20\,MeV by a power-law with index 2.24, as shown in Fig.\,\ref{fig:InputSpectra}.
The Crab simulation was performed in eight separate energy ranges. For each bin, a simulation with 500k pre-trigger events was performed, resulting in a total of 2.85\,M events with a distribution shown in Table \ref{tab:CrabComposition}. The source was set up as a far-field point source with an incident angle $\theta = 30^\circ$ . MEGAlib was tasked with reconstructing Compton and Pair events and eliminating events from minimum ionizing particle tracks (MIPs). It applies a  cascade of reconstruction algorithms with increasing  computational complexity to each event. If at any stage the algorithm is successful, the chain stops and the event is classified accordingly. The MIP-search tests the hit-pattern for linearity by applying a linear $\chi^2$ fit. The next stage is the Pair-reconstruction. Here a pattern-recognition algorithm is searching for a creation vertex. If successful, the electron and positron tracks are reconstructed by applying a Pearson correlation test. The Compton reconstruction is separated into two stages. First, the recoil electron track is extracted and reconstructed based on a Spearman-Rank test. Then the Compton-sequence is reconstructed by minimizing the $\chi^2$ of the difference between the kinematic scatter angle obtained via Compton's formula and the geometric scatter angle derived from the hits' locations. Details about the implementation in MEGAlib can be found in\cite{ZogDiss}.

\begin{table}[hbt!]
\caption{Upper limit of the simulation bins, medians and contributing events per bin. The minimum photon energy was set to 60\,keV. The total size of the dataset is 2.85 million events.}
    \centering
    \begin{tabular}{c|c|c|c|c|c|c|c|c|c}
      E$_{max}$[MeV] &  0.13 & 0.26 & 0.45 & 1.1 & 2.4 & 4.7 & 10 & 20 \\
        \hline      
        Median [MeV] & 0.07 & 0.16 & 0.31 & 0.56 & 1.35 & 2.85 & 5.72 & 12.03 \\
        \hline
        Events & 106k & 289k & 378k & 415k & 422k & 422k & 413k & 407k
    \end{tabular}
    \label{tab:CrabComposition}
\end{table}

We use the result of the reconstruction as prediction of the event type of the baseline classification.
We considered an event tagged as Compton when the reconstruction succeeded in Compton reconstruction. The event was considered tagged as pair, when the pair reconstruction accepted the event. An event was considered tagged as non-reconstructable when it was either tagged  as MIP or all reconstruction algorithms rejected the event.

\begin{figure}[hbt!]
    \centering
    \includegraphics[width=0.95\textwidth,trim={40pt 10pt 60pt 50pt},clip]
    {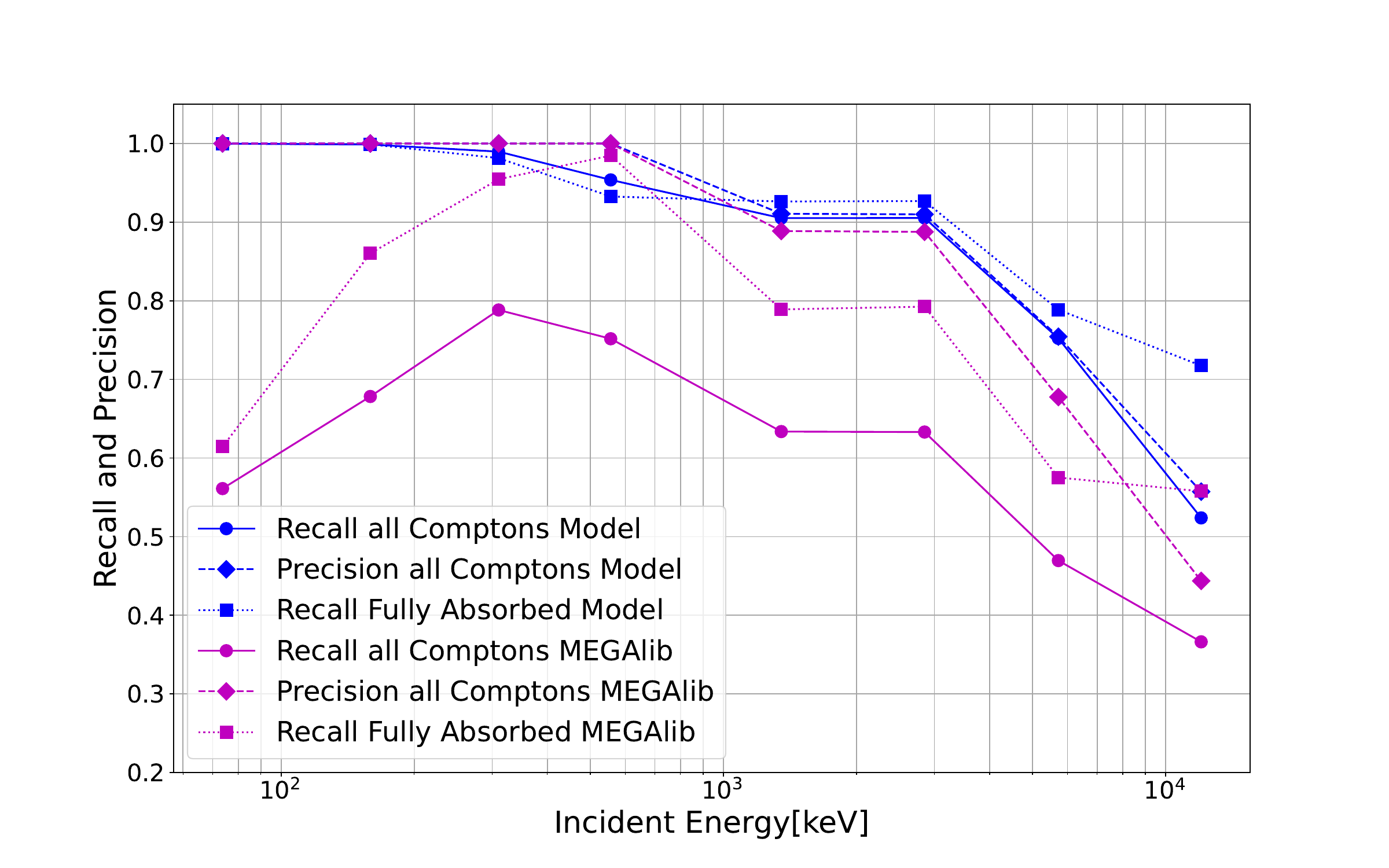}
    \caption{Precision and Recall for Compton events as function of energy for the Crab-source dataset.}
    \label{fig:precRecComp}
\end{figure}

The performance on the Compton dataset is shown in Figure \ref{fig:precRecComp}. The default reconstruction tags up to 80\% of all Compton events correctly at its peak around 550\,keV and then drops off steadily at higher energies. The classification by our model outperforms the default classification, giving significantly higher recall values for each energy bin of the crab spectrum on the full Compton dataset. The precision on the low-energy bins is nearly perfect in both classifications with a slight edge at higher energies for the model classification. An important subset constitute fully absorbed Compton events, which have the highest likelihood to be correctly reconstructed and therefore contain the proper directional information of the event. 
Here the model classification outperforms the default classification in every but one bin at 550\,keV (93\,\% vs 98\,\%) by up to 38 percent-points.

\begin{figure}[hbt!]
    \centering
    \includegraphics[width=0.95\textwidth,trim={40pt 10pt 60pt 50pt},clip]
    {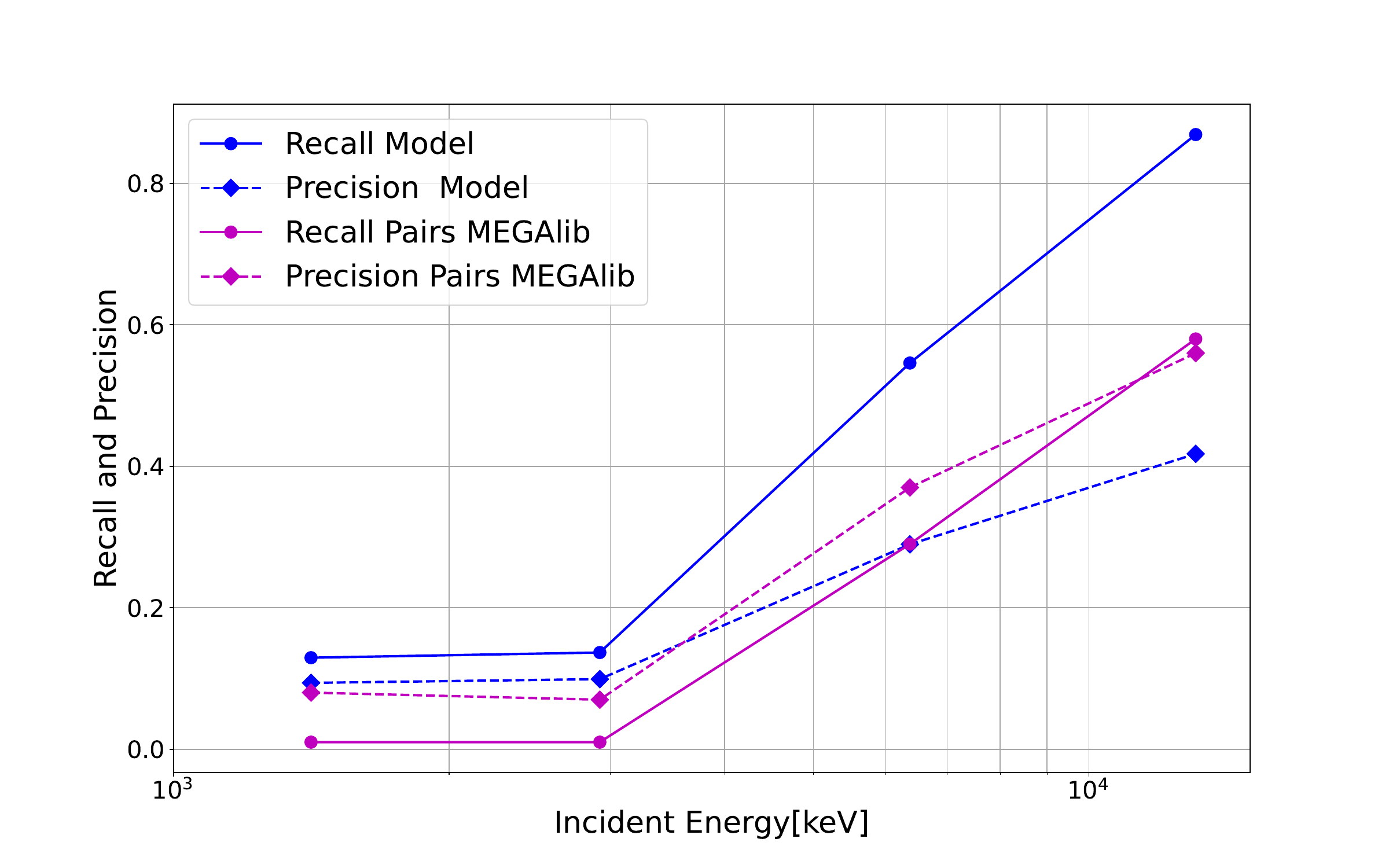}
    \caption{Precision and Recall for Pair events as function of energy for the Crab-source dataset.}
    \label{fig:precRecPair}
\end{figure}

\begin{figure}[hbt!]
    \centering
    \includegraphics[width=0.95\textwidth,trim={40pt 10pt 60pt 50pt},clip]
    {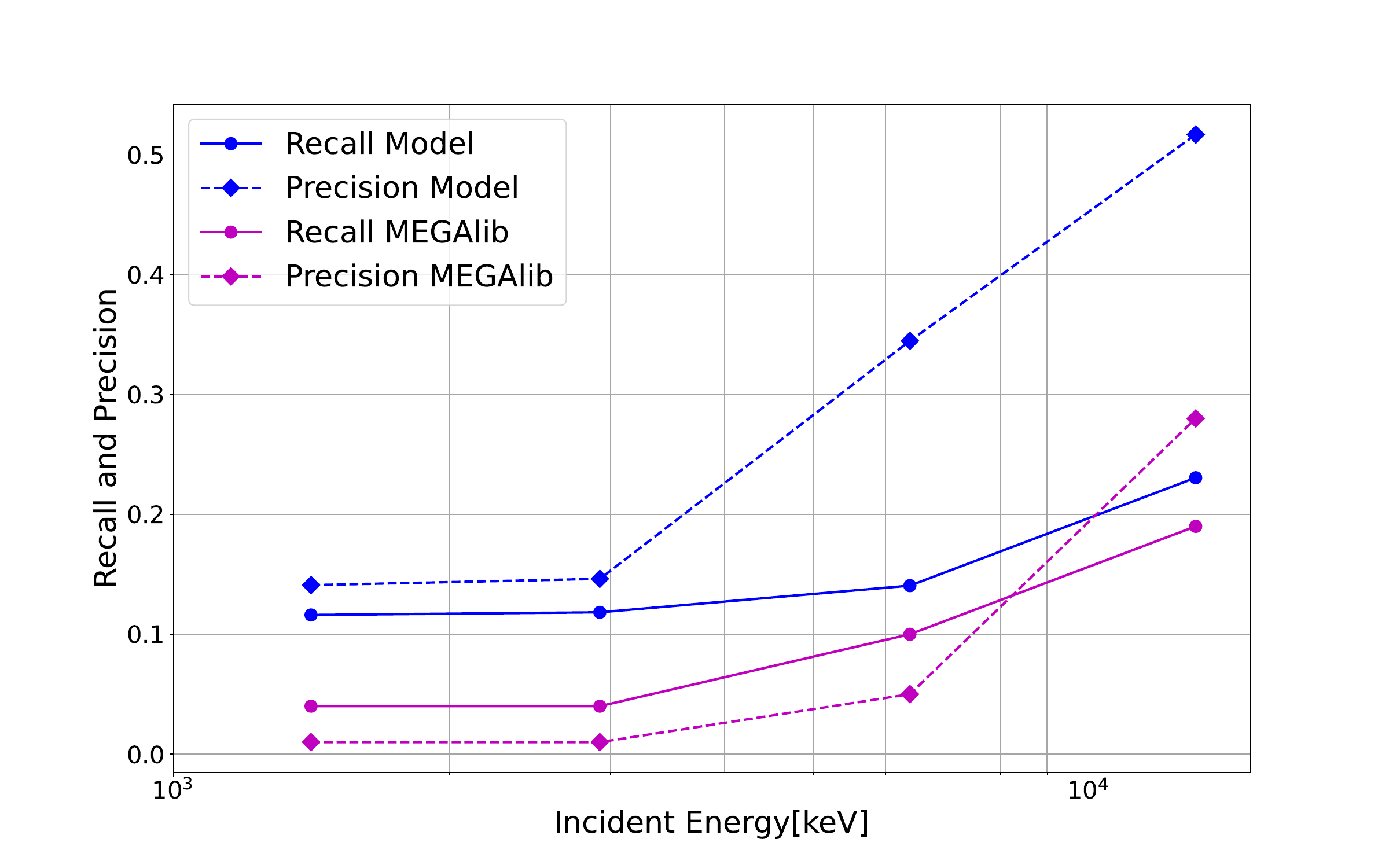}
    \caption{Precision and Recall for non-reconstructable events as function of energy for the Crab-source dataset.}
    \label{fig:precRecTrash}
\end{figure}
The performance on D1 and D2 pairs is depicted in Figures \ref{fig:precRecPair} and \ref{fig:precRecTrash} . At energies above the pair threshold and below 5\,MeV, both the model and the default classification show very little discrimination power. Nonetheless, the model's Recall on D2-first pairs shows an improvement over the default case. The D1-pair Recall shows large improvements over the default classification, starting to recognize D1-pairs  reliably at energies larger than 5\,MeV and a very good class-coverage of 87\,\% at the highest simulated energies. The Precision of the model classifications on D1 and D2-pairs shows a slightly worse purity of the pair datasets after classification than in the default case, which can be explained by the relatively large amount of incorrectly tagged D2-pairs which have been tagged as D1-pair and vice-versa (see Figs.\,\ref{fig:comptonConf}, \ref{fig:pairConf} and \ref{fig:trashConf}).

\begin{figure}[hbt!]
    \centering
    %[trim={left bottom right top},clip]
    \includegraphics[width=0.95\textwidth,trim={40pt 10pt 60pt 50pt},clip]{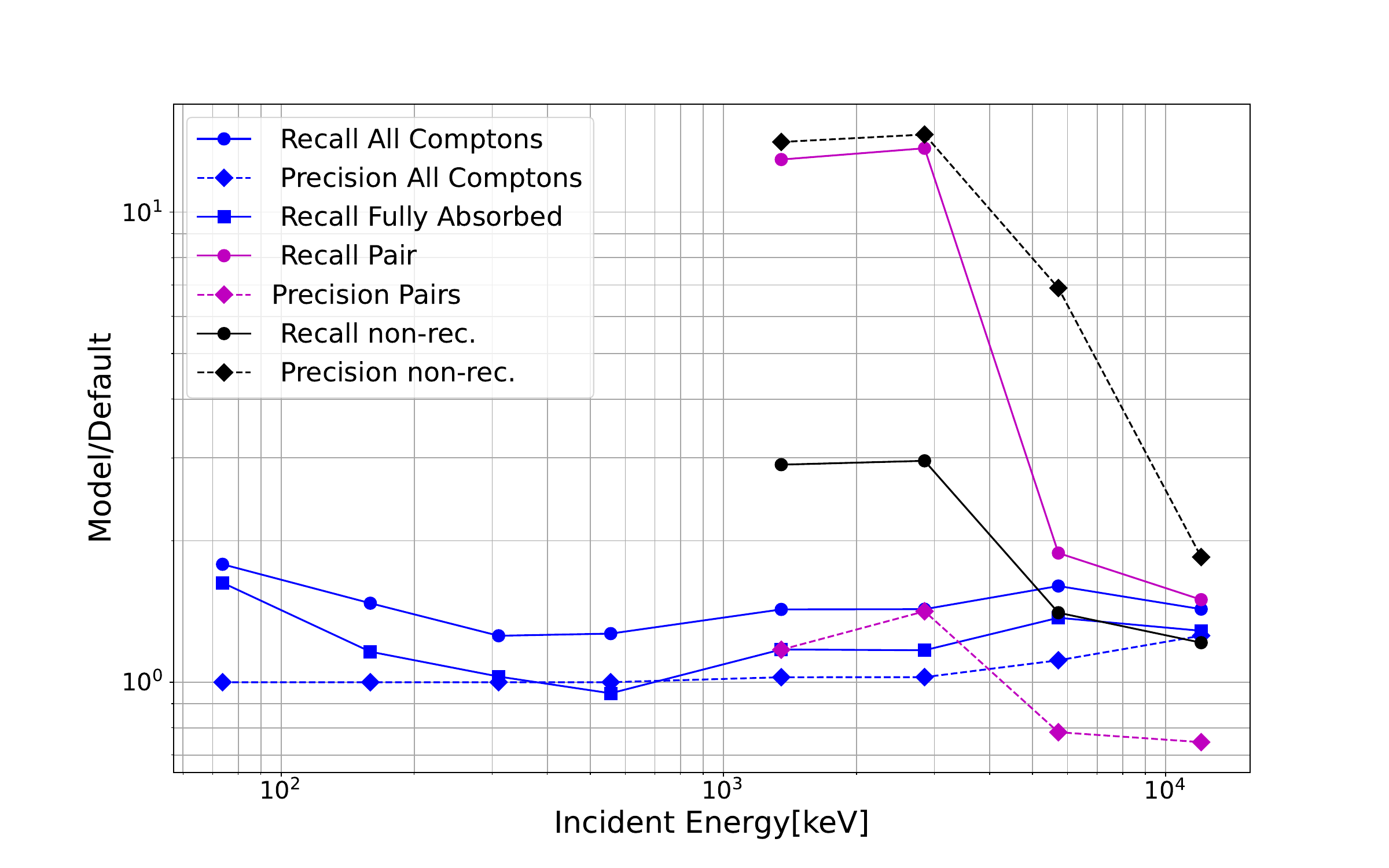}
    \caption{Relative improvement of the model classification vs the default reconstruction as function of energy for the Crab-source only dataset.}
    \label{fig:modelRatio}
\end{figure}

Figure \ref{fig:modelRatio} shows the relative improvement of the model classification over the default case, given as the ratio of model performance relative to the default. Data points above 1 indicate improvement, below 1 indicate worse performance than the default. The recall on all Comptons, D1 and D2 pairs shows an overall improvement of around 25\% to 60\% over the default classification. In the low energy bins, the improvement ratio for D1-pairs reaches beyond a factor 10 while maintaining an improvement ratio of greater than 1.5 for high energies. Even though the recall on D2-pairs in this energy range is still sub-optimal, the model's performance is an improvement over the default classification over the whole inspected energy range, improving the class coverage by a factor 3 for lower energies to a factor of 1.3 at the higher energies. The observed behaviour of the model classification is consistent with the expectations from the omni-directional test set.

\section{Summary and Outlook}
The primary goal of this work was to explore the possibility to detect event types and event topologies with CNN-driven models. The successful model presented here is capable of distinguishing different event types and undesired event topologies, as shown on an omni-directional test set as well as on a simulated Crab measurement.   

This model was trained to distinguish Compton, pair and non-reconstructable events. The latter consisted of signal photons with a non-reconstructable event topology and events caused by leptons, like cosmic and albedo electrons and positrons.  Of course there are even more types of background events, like events resulting from delayed decays of activated materials, and events caused by background hadrons. These components will be the subject of a follow-up study.

The presented model has a computational overhead of about 0.9 GOPS per event. which fits well into the computational envelope of recently developed embedded systems. An architecture like the presented one is capable of processing even larger rates for smaller inputs. Studies are under way to test these  embedded systems, making an online-approach viable, e.g., for CubeSat scenarios.

One particular aspect of this architecture remains unexplored. Since SPP enables the use of dynamic image sizes, it is possible to decrease the number of necessary operations per event even further by cropping the input images to a smaller size. For computational estimates, the average image size then becomes  a variable of the measurement environment. Such a study would make sense once a mission is funded and various orbit scenarios have to be explored.

%\begin{acknowledgement}
\subsubsection*{Acknowledgement}
 Jan Peter Lommler acknowledges support by DLR grant 50OO2218.
%\end{acknowledgement}

\clearpage
%\newpage

%\newpage

\bibliography{sn-bibliography}

\nocite{*}

\end{document}